\documentclass[aps,prl,reprint,superscriptaddress]{revtex4-1}

\usepackage{graphicx}  
\usepackage{bm}        
\usepackage{amsmath}   
\usepackage{amssymb}   
\usepackage{epstopdf}  

\usepackage{hyperref}

\usepackage{color}

\newcommand{\phdagg}{{\phantom{\dagger}}}
\newcommand{\ph}{{\phantom{0}}}
\newcommand{\ket}[1]{|{#1}\rangle}

\newcommand{\braket}[2]{\langle{#1}|{#2}\rangle}

\newcommand{\e}{{\mathrm{e}}}

\begin{document}

\title{Braiding errors in interacting Majorana quantum wires}



\author{Michael Sekania}
\affiliation{Institute for Theoretical Physics and Astrophysics,
Julius-Maximilian University of W\"urzburg,
Am Hubland, 97074 W\"urzburg,
Germany}
\affiliation{Center for Electronic Correlations and Magnetism,
Institute of Physics, University of Augsburg, 86135 Augsburg,
Germany}
\affiliation{Andronikashvili Institute of Physics,
Tamarashvili~6, 0177 Tbilisi,
Georgia}

\author{Stephan Plugge}
\affiliation{Institut f\"ur Theoretische Physik,
  Heinrich-Heine-Universit\"at, 40225 D\"usseldorf, Germany}

\author{Martin Greiter}
\affiliation{Institute for Theoretical Physics and Astrophysics,
Julius-Maximilian University of W\"urzburg,
Am Hubland, 97074 W\"urzburg,
Germany}

\author{Ronny Thomale}
\affiliation{Institute for Theoretical Physics and Astrophysics,
Julius-Maximilian University of W\"urzburg,
Am Hubland, 97074 W\"urzburg,
Germany}

\author{Peter Schmitteckert}
\affiliation{Institute for Theoretical Physics and Astrophysics,
Julius-Maximilian University of W\"urzburg,
Am Hubland, 97074 W\"urzburg,
Germany}
\affiliation{Institut f\"ur Theoretische Physik,
  Heinrich-Heine-Universit\"at, 40225 D\"usseldorf, Germany}


\date{\today}

\begin{abstract}
Avenues of Majorana bound states (MBSs) have become one of the primary directions towards a possible realization of topological quantum computation. 
For a Y-junction of Kitaev quantum wires, we numerically investigate the braiding of MBSs while considering the full quasi-particle background.
The two central sources of braiding errors are found to be the fidelity loss due to the incomplete adiabaticity of the braiding operation as well as the hybridization of the MBS.
The explicit extraction of the braiding phase in the low-energy
Majorana sector from the full many-particle Hilbert space allows us to
analyze the breakdown of the independent-particle picture of Majorana
braiding.
Furthermore, we find nearest-neighbor interactions to significantly affect the braiding performance to the better or worse, depending on the sign and magnitude of the coupling.
\end{abstract}


\maketitle

\paragraph{Introduction.}
Topological quantum states of matter have become one of the most vibrant fields of contemporary condensed matter physics.
One subbranch thereof, the search for realizations of topological quantum computation, has decisively fueled the interest both from theory and experiment to address manifold scenarios of this kind.
Among them, Majorana bound states (MBS) have been attracting pivotal
attention in recent years {\cite{Alicea2012,LeijnseFlensberg2012,Beenakker2013,DasSarma2015}.
While Majorana-type quasiparticles have been
previously addressed in the context of the fractional quantum Hall effect \cite{Moore1991,Read2000},
unconventional superconductivity \cite{Read2000,Kitaev2001,Ivanov2001},
and spin liquids \cite{Kitaev2006,Greiter2009},
the proposal by Fu and Kane \cite{FuKane2008} to employ conventional superconducting proximity effect to stabilize MBS in 
vortex cores at the surface of a topological insulator unleashed a
remarkable body of research that has brought the detection and
manipulation of MBS much closer to reality. 
After crucial progress towards simpler realizations of proximity-induced topological superconducting phases \cite{Lutchyn2010,Oreg2010,NadjPerge2013},
several experimental groups have reported increasingly compelling evidence for the observation of Majorana zero modes
in semiconductor nanowires and iron atomic chains that are coupled to a bulk superconductor \cite{Mourik2012,Rokhinson2012,Das2012,Deng2012,Churchill2013,Yazdani2014,Albrecht2016,Deng2016,Pawlak2016,Feldman2017,Zhang2016}.
The unambiguous detection of MBS, however, has so far remained elusive, and parity-changing quasiparticle poisoning through single-particle tunneling into the wire or chain likewise constitutes a major challenge. 
One crucial experimental finding in favor of the existence of Majorana zero modes would be a braiding experiment \cite{Alicea2011,vanHeck2012,Li2016,Aasen2016}, revealing their non-trivial braiding statistics, and showing the path to more complex multi-MBS protocols.

A plethora of approaches is currently underway towards the realization of topological quantum computation in networks of semiconductor nanowires proximity-coupled to a superconductor \cite{Hyart2013,Aasen2016,Landau2016,Plugge2016,Plugge2017,Vijay2016,StationQ2016}.
Alternative routes include the aforementioned proximitized topological-insulator surfaces \cite{FuKane2008,Vijay2015,Vijay2015b}, and atomic chains \cite{NadjPerge2013,Li2016} or cold-atom setups \cite{Kraus2013,Laflamme2014} where the originally envisioned lattice realizations of topological superconductivity \cite{Kitaev2001} are more directly accessible.
Given the amount of different obstacles still to overcome, it is hard to predict which direction will succeed.
At the current stage of the field, it thus appears worthwhile following up on several of these directions at the same time: New results on any given approach will offer guidance for inevitable refinements of the others.

In this paper, we numerically analyze the braiding process of MBSs in a minimal model of an interacting wire network.
We define a Y-junction of three Kitaev chains~\cite{Kitaev2001} in which we investigate the controlled time-evolution of a single pair of MBSs as the most elementary braiding operation.
We employ time-dependent on-site potentials to locally drive the system in and out of the topologically non-trivial phase, which hence allows us to spatially move the MBSs \cite{Alicea2011}.
The potential-manipulation protocol is analyzed to yield an optimal performance as to minimize non-adiabatic effects.
The small system sizes accessible through exact diagonalization allow us to systematically identify the different sources of braiding errors.
Since we keep the full Hilbert space during the braiding operation, we can correlate the taken protocol with the fidelity of the MBS sector in which we operate.
In addition, this allows us to extract explicitly the geometric
exchange ``braiding'' phase acquired during the execution of the braid, directly encoding the non-Abelian statistics of MBSs.
By varying the superconducting wire bulk gap, we then find how the
localization length of the individual Majorana zero modes as well as
the resulting MBSs hybridization affect the braiding operation.
Finally, interactions in the wires, modeled by a nearest-neighbor density-density coupling \cite{Loss2011,Stoudenmire2011,Sela2011}, are found to predominantly affect the braiding operation through its impact on the wire bulk gap.
Depending on the non-interacting gap versus hopping strength, for weak attractive (repulsive) interaction, the braiding operation is more stable due to an increase (decrease) of the effective superconducting gap and reduction of the MBS localization length. 
For strong attractive or repulsive nearest-neighbor coupling, the interactions have a negative impact on the braiding performance, where ultimately the topological phase becomes inaccessible altogether~\cite{Loss2011,Stoudenmire2011,Sela2011}.

\paragraph{Kitaev chain.}
The (interacting) Kitaev model is the most elementary system  exhibiting MBSs \cite{Kitaev2001,Loss2011}. It is given by the Hamiltonian \cite{HasslerSchuricht2012,ThomaleRachelSchmitteckert2013}
\begin{eqnarray}
 \label{eq:KitaevChain}
 H &=& \sum\limits_{i=1}^{L - 1}
           (  J      c^\dagger_i c^\phdagg_{i+1}
            + \Delta c^\dagger_i c^\dagger_{i+1}
            + \text{H.c.})
   + \sum\limits_{i=1}^{L}
           \mu_i^\phdagg   n_i \nonumber \\
&& \qquad +~\sum_{i=1}^{L-1}V n_i n_{i+1}
\end{eqnarray}
where $c^\dagger_i$ ($c^\phdagg_i$) is a creation (annihilation) operator of the spinless fermion on site $i$ and $n_i=c^\dagger_i c^\phdagg_i$ is the density operator.
$J$ denotes the nearest neighbor hopping amplitude, $\Delta=|\Delta| \e^{i\varphi}_\ph $ the $p$-wave superconducting pairing amplitude with phase $\varphi$, $V$ the strength of the nearest neighbor interaction, and $\mu_i$ the site-dependent potential.
The superconducting pairing term breaks $U(1)$ fermion number symmetry down to $\mathbb{Z}_2$ (parity conservation), where the two parity sectors will be denoted even $(e)$ and odd $(o)$ below.
Unless stated otherwise, we initially constrain ourselves to the non-interacting limit $V=0$.
The Kitaev chain features a topological trivial and non-trivial phase  for $|\mu| > 2J$ and $|\mu| \leqslant 2J$, respectively, so long as $|\Delta| > 0$ \cite{Kitaev2001,Alicea2012}.
For the ideal parameter set  $\mu=0$, $|\Delta| = J$ and hard wall boundary conditions, hence residing in the non-trivial regime, Eq.~\eqref{eq:KitaevChain} takes a particularly simple form in terms of a Majorana fermion representation 
$H=-J\sum_{j=1}^{L-1}i\gamma^\phdagg_{2j}\gamma^\phdagg_{2j+1}$,
where 
$\gamma^\phdagg_{2j-1} =  \e^{i\varphi/2}_\ph c^\phdagg_j + \e^{-i\varphi/2}_\ph c^\dagger_j$
and
$\gamma^\phdagg_{2j}   = (\e^{i\varphi/2}_\ph c^\phdagg_j - \e^{-i\varphi/2}_\ph c^\dagger_j)/i$
are real Majorana operators $\gamma^\dagger_i = \gamma^\phdagg_i$ obeying a Clifford algebra, $\{\gamma^\phdagg_i,\gamma^\phdagg_j\}=2\delta^\phdagg_{ij}$.
We observe that the first and last Majorana modes, residing on the first and last site, respectively, are decoupled from the rest of the fermionic chain.
The corresponding Majorana operators,
$\gamma^\phdagg_1 \equiv \gamma^\phdagg_\mathrm{L}$ and $\gamma^\phdagg_{2L}  \equiv \gamma^\phdagg_\mathrm{R}$,
do not appear in the Hamiltonian, $[H,\gamma^\phdagg_\mathrm{L}]=[H,\gamma^\phdagg_\mathrm{R}]=0$,
and form a fermionic zero-energy mode comprised by two individual Majorana (zero) modes localized on single sites at each end of the chain.
After introducing a non-local fermionic operator $f = \frac{1}{2}(\gamma_\mathrm{L} + i \gamma_\mathrm{R})$,
these zero-energy states can be identified with MBS eigenstates $f\ket{0} = 0$ and $f_\ph^\dagger\ket{0} = \ket{1}$,
which reside in the middle of the effective bulk gap $4|\Delta|/J$.
The states $\ket{0}$ and $\ket{1}$ also coincide with the parity eigenstates of the full fermionic chain,
and can be associated to the ground state wavefunctions of the Kitaev Hamiltonian \eqref{eq:KitaevChain} in the even ($e$) and odd ($o$) parity sector, respectively~\cite{mvr}.

Away from this particular parameter set but still within the
non-trivial regime, i.e. for $|\Delta| < J$ or $0<\mu<2J$,
the Majorana zero modes develop an exponential tail extending into
the bulk, yielding a small energy split of the two eigenstates
previously located at zero energy (in different parity sectors).
This splitting energy, usually referred to as MBS hybridization energy, decays exponentially with the distance between MBSs
on a length-scale governed by $\Delta$. 
The protection offered by the bulk energy gap and spatial isolation of
MBSs are the main stimuli for $\ket{0}$ and $\ket{1}$ to serve as a basis for the sought-after topological qubit.
As we show below, targeting the entire Hilbert space -- and not only a subspace of sub-gap Majorana states in terms of an effective low-energy theory as conventionally assumed~\cite{Alicea2011,Aasen2016,Cheng2011,Clarke2011,Halperin2012,Hyart2013,Karzig2013,Scheurer2013,Karzig2015a,Karzig2015b,Kraus2013,Laflamme2014,Landau2016,Li2016,Plugge2016,Plugge2017,StationQ2016,Vijay2016,vanHeck2012,Vijay2015,Vijay2015b}
-- is essential when manipulations such as braiding are performed on finite time scales (non-adiabatically) and in finite-size systems. Preceding works undertook steps in this direction \cite{Amorim2015,Pedrocchi2015a,Knapp2016}, but attempted to ground the errors in an effective low-energy (Majorana) picture.
%

In the following,
via a sequence of predefined exact manipulations of potenitals $\mu_i$ of Hamiltonian \eqref{eq:KitaevChain},
we perform a braiding operation imposed on an initially prepared set of MBSs \cite{Alicea2011}. Such manipulations employ unitary but in general non-adiabatic dynamics.
This implementation requires a way to
(i) dislocate MBSs without destroying them or creating new ones,
(ii) realize a system geometry that allows to exchange Majorana modes while keeping their overlap exponentially small during the entire braiding process,
and (iii) provide an exchange process that gives fidelity, taken as the overlap between initial and final states, close to unity.

At the elementary level of our description, the goal is not to define the optimal exchange protocol applied to a realistic setup for topological quantum computation (though we made sure to select a favorable implementation among several ``simple'' test routines).
Rather, for the simplest braiding operation possible, we intend to identify the time scale and conditions under which it is feasible to perform a braiding with the desired fidelity and braid-phase accuracy.
For an analysis of tailor-made braid protocols and possible drawbacks or limitations thereof, based on effective models of the low-energy (Majorana) sector of the system, cf. Refs. \cite{Karzig2015a,Karzig2015b,Knapp2016}.

\begin{figure}
 \includegraphics[width=\columnwidth]{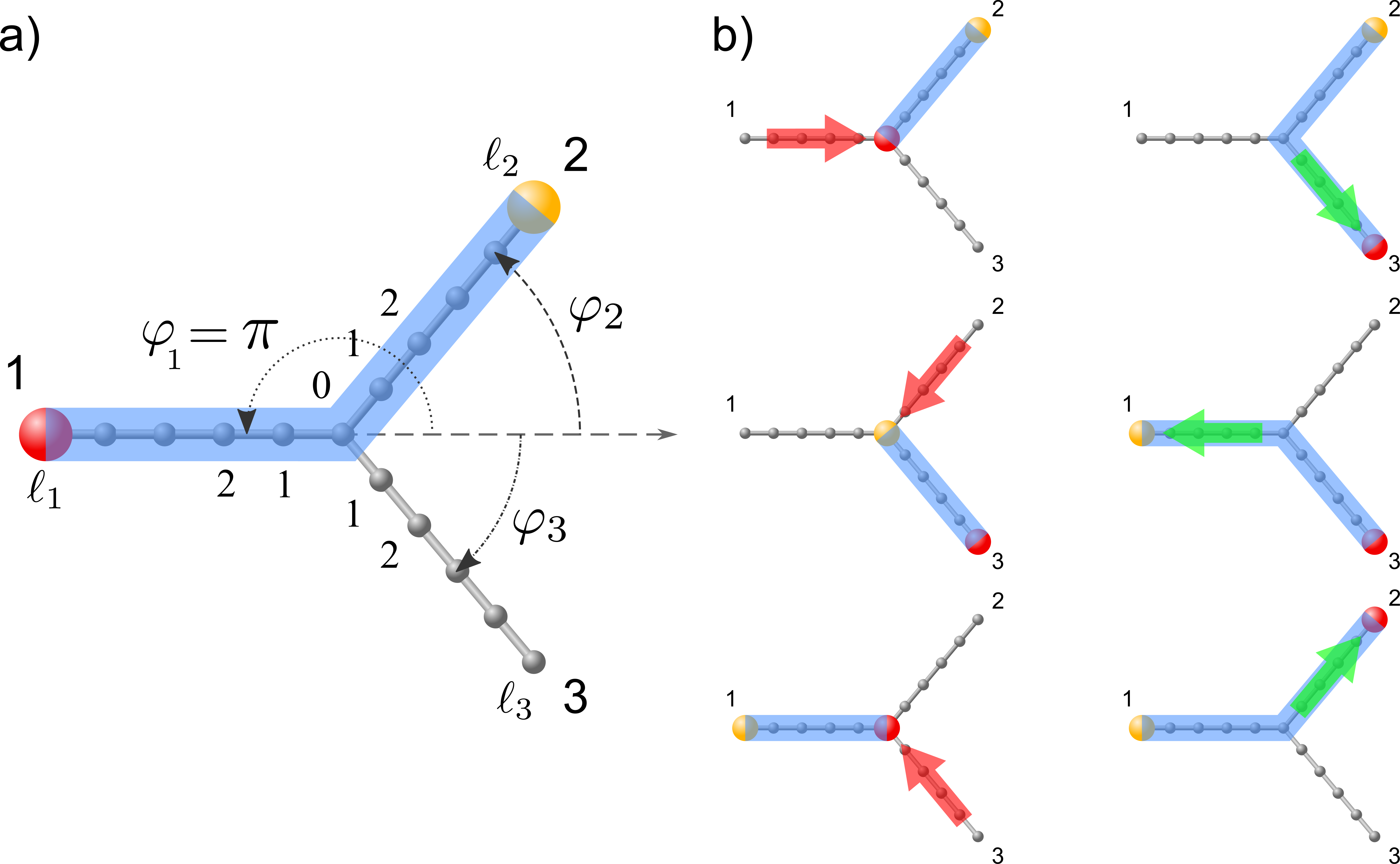}
 \caption{\label{fig_1}
          Y-junction of three Kitaev chains and the braiding protocol considered here.
          MBSs, depicted by red and yellow spheres, reside at the ends of segments with topological non-trivial phase (
          highlighted by blue tubes).
          a)~The initial state. Geometric angles $\varphi_i$ also correspond to the superconducting pairing phases of the  respective junction legs.
          b)~The exchange ``braiding'' protocol (left-to-right and top-to-bottom), as in Ref. \cite{Alicea2011}.
          Red (green) arrows indicate displacements of MBSs by contraction (extension) of the topological chain segment, implemented by locally ramping up (down) the local potential over the critical value $\mu_c=2J$.
          At the end of the braid protocol (right picture, bottom row), the same chain segments as in the initial state reside in the topological phase, but the MBSs have been exchanged.
         }
\end{figure}

\paragraph{Y-junction.}
The minimal geometry required for the exchange ``braiding'' of MBSs is a junction with at least three legs (tri-junction) \cite{Alicea2011,Alicea2012}. 
It can be formed by connecting three Kitaev chains (later referred to as junction legs) to an additional connecting junction site at $i=0$ \footnote{The site indices in each junction leg are taken as increasing from the central tri-junction site to the free edge.}, which otherwise witnesses parameters sets identical to the respective adjacent leg:
\begin{eqnarray}
 H^\phdagg_\mathrm{Y} =
       \mu_0^\phdagg c^\dagger_0 c^\phdagg_0
   &+& \sum\limits_{n=1}^{3}
         \left(
             J^{(n)}_\ph      c^{\dagger}_0 c^{(n)}_{1}
           + \Delta^{(n)}_\ph c^{\dagger}_0 c^{\dagger,(n)}_{1}
           + \text{H.c.}
         \right)                   \nonumber \\
   &+& \sum\limits_{n=1}^{3} V n_0 n_{1}^{(n)}
      +\sum\limits_{n=1}^{3} H^{(n)}.  \label{eq:y_junction}
\end{eqnarray}
Here $H^{(n)}$ for each leg 
corresponds to the Kitaev Hamiltonian \eqref{eq:KitaevChain} with
parameters $J^{(n)}$, $\Delta^{(n)}$, $V^{(n)}$, and $\mu^{(n)}_i$.
In general, one may choose different $p$-wave pairing terms and hopping amplitudes in each leg of the Y-junction.
For simplicity, we here assume that $J^{(n)} = 1$ (thus setting the reference scale in plots and equations below), $\Delta^{(1)} = -\Delta$, $\Delta^{(2)} =  \Delta\,e^{i\varphi_2}$, and
$\Delta^{(3)} =  \Delta\,\e^{i\varphi_3}$,
with a real pairing amplitude $\Delta > 0$.
The superconducting pairing phase $\varphi_n$ is chosen to coincide with the corresponding geometric angle of the $n$-th leg of the Y-junction, cf. Fig.~\ref{fig_1}.
This type of triplet pairing could, for example, be realized in a Y-junction placed in proximity to a chiral $p$-wave superconductor \cite{Halperin2012}.
The connection of two Kitaev chain legs, both of which are in the
topological (trivial) phase, produces a single segment residing in
the topological (trivial) phase;
unless the mutual phase difference between the pairing terms of these chains equals exactly $\varphi_i-\varphi_j = \pi$,
MBSs on the connecting site fuse,
and the entire segment ends up with the two left-over MBS residing on the free ends of the new, larger topological segment \cite{Alicea2011}.
If the mutual phase, however, is exactly $\pi$,
MBSs on the connecting site do not fuse and the joint segment
maintains all four MBSs (one from each end of the initial chains) -- a case we intentionally avoid in the following.

By ramping up the on-site potentials over the critical value of
$\mu_c=2J$ starting from one end of the topological segment (i.e., at
the location of a Majorana zero mode), locally one drives 
the system into the trivial phase. 
As the ramp of potentials is successively carried out along the chain, one thus continuously displaces the MBS alongside with the trivial-topological domain wall.
This action can be undone by again lowering the local potentials below the critical value, and switching the chosen sites back from the trivial to topological phase. Finally, concerting such
ramping procedures in sequences across all three junction legs, one 
can create the desired Majorana zero mode exchange or braiding operation \cite{Alicea2011}.
This process is schematically depicted in Fig.~\ref{fig_1}.
We will only study Y-junctions with legs of equal length $L^{(n)}=\ell$, and take $\varphi_2=-\varphi_3=\varphi$.
The initially homogeneous potentials in the junction legs are $\mu^{(1)}=\mu^{(2)}=0$ and $\mu^{(3)} = 2\mu_c$, meaning that legs $1$ and $2$ reside in the topological, and leg $3$ in the trivial phase. In passing we note that since the time evolution of an initial state prepared like this fixes a time
direction, there is a subtle time-reversal symmetry breaking appearing in the braiding setup as seen by reversing the sign of $\varphi$ in the time evolution.
We thus only consider $\varphi >0$ in the following.

\paragraph{Ramping protocol.}
We investigated several ramping protocols for the successive time variation of the on-site potentials $\mu_i^{(n)} (t)$.
The most stable results were found by employing a sine-squared ramp defined as
\begin{align}
 \label{eq:ramp}
 \mu^{(n)}_j(\tau) = 2\mu_c\,m\left(\frac{\tau}{T}[1 + \alpha (\ell - 1)] - \alpha (\ell - j)\right)~,
\end{align}
where $\tau \in [0,T]$ denotes time during the respective ramp-up protocol step in Fig.~\ref{fig_1}, and we used the scalar function
\begin{align}
 \label{eq:sin_ramp}
 m(q) &= \sin^2\left(\frac{\pi}{2}r(q) \right)\,.
\end{align}
Here, $r(q)$ is a linear ramp of unit height and duration
\begin{align}
 r(q) &= \min\left[\max(q,0),\,1 \right]  
       = \left\{
          \begin{array}{cll}
           0 & & q<0\\[5pt]
           q & & 0\leqslant q \leqslant 1\\[5pt]
           1 & & q > 1
          \end{array}
         \right.. \nonumber
\end{align}
We also show results obtained by the simpler, linear ``guillotine'' ramp, $m(q) = r(q)$,
where the smoothening by the sine-squared function in Eq.~\eqref{eq:sin_ramp} is switched off.
Both procedures raise the on-site potentials in the $n$th leg, containing $\ell$ sites, from $\mu^{(n)}_\text{init} = 0$ to $\mu^{(n)}_\text{f} =2\mu_c=4J$ within the time period $T$.
The time required to lift the on-site potential on each individual site is $T/(1 + \alpha(\ell-1))$, and the ramping is delayed by
$\alpha T/(1 + \alpha(\ell-1))$ between consecutive sites.
For a guillotine ramp, the modulations of local potentials are reminiscent of a guillotine knife passing a rectangular window (hence the name), and $\alpha$ is the inclination of the knife.

The delay coefficient (inclination) can be varied between $\alpha = 0$ and $\alpha = 1$, corresponding to a simultaneous ramp of all sites or a consecutive site-by-site ramping, respectively.
In the following, we set $\alpha = 0.025$ and vary the time step $T$, where $\alpha^{-1}, T \gg \ell$ at all times, i.e., a small ramp delay (inclination) and slow ramp protocol.
Finally, the ramp-down is implemented as an exact time-reverse of the ramp-up protocol, with $\tau \to T-\tau$ in Eq.~\eqref{eq:ramp}.
The ramping (and hence braiding) protocol is accomplished numerically by considering the piece-wise constant, time-independent Hamiltonian
\begin{eqnarray}
 \label{eq:time_evol}
 \ket{\psi(t + T)} &=& \mathcal{T} \e^{-i \int_{t}^{t + T}H^\phdagg_\mathrm{Y}(t)\,\mathrm{d}t} \ket{\psi(t)}  \nonumber \\
                   &=& \left(\mathcal{T} \prod_{j = 0}^{N - 1} \e^{-i
                       H^\phdagg_\mathrm{Y}\left(t + j\Delta
                         t\right)\,\Delta t} \right) \ket{\psi(t)}\, ,
\end{eqnarray}
where $\mathcal{T}$ is the time ordering operator,
and $H^\phdagg_\mathrm{Y}(t)$ corresponds to~\eqref{eq:y_junction}
with time-dependent local potential $\mu^{(n)}_j \rightarrow \mu^{(n)}_j(t)$ given in \eqref{eq:ramp}.
The time-step discretization $\Delta t / T \ll 1$ governs the accuracy of approximation of the continuous time-dependent Hamiltonian by a piece-wise constant one, and is chosen sufficiently small against all energy scales of the problem, yielding a total of $N=T/\Delta t$ steps per braiding sweep in Fig.~\ref{fig_1}.
As such, the outlined braiding procedure demands a total (final) braid time $t_{\text{f}}=6T$, and performs a complete cyclic evolution in the parameter space of the Hamiltonian.

As braiding errors are detected, there are several parameters that can be used to control and alter the braiding process. The ramping speed can be controlled by the time period $T$ required for each step of the protocol, and by the ramp delay/inclination $\alpha$.
Since the latter also determines the steepness of the trivial-topological domain wall propagating through the chain, generally, for steeper inclinations, a slower braid velocity $\sim \ell/T$ with larger time period $T$ is required \cite{Karzig2013,Scheurer2013}.
Furthermore, the rate $\Delta t/T$ above determines the smoothness of the ramping potential. In the following, we take $\Delta t /T \ll 1$ small against all relevant energy scales and simulate an effectively smooth ramp.
%

\paragraph{Fidelity and geometric braiding phase.}
In order to achieve an ideal braiding process, it is necessary (but not sufficient) that the fidelity between the initial and obtained final state equals unity.
Since numerical simulations of the full Hilbert space are
costy in general, and sampling the fidelity for many input states
-- in particular if one is interested in longer braid sequences --
is not feasible, we propose to analyze the topological character and protection of braid operations 
for the ground states of the two parity sectors.
For this purpose, the braiding statistics are directly encoded in the relative geometric phase factor (the braid phase) between final and initial states in the even and odd parity sector.
Braiding errors then manifest both in deviations of the braid fidelity from unity and of the braiding phase from $\pi/2$.
Both quantities are accessible by considering input states $\ket{0}$ and $\ket{1}$ of the topological qubit, chosen as the ground states of $H_{\mathrm{Y}}(t_0=0)$ \eqref{eq:y_junction} in the respective total-parity sector.
(Note that transitions between even $(e)$ and odd $(o)$ total-parity states are forbidden under unitary time evolution \eqref{eq:time_evol} with a parity-conserving Hamiltonian \eqref{eq:y_junction}.)

In order to quantify the fidelity loss during the braiding, we introduce the loss function
\begin{equation}
 \label{eq:lost_fidelity}
 w_\mathrm{loss}(t_\mathrm{f}) = 1 - F(t_\mathrm{f})^2 = 1 -
 |\braket{\psi(t_\mathrm{f})}{\psi(t_0)}|^2\, ,
\end{equation}
where $F(t_\mathrm{f}) = |\braket{\psi(t_\mathrm{f})}{\psi(t_0)}|$ is the fidelity for pure states~\cite{nielsen00}, taken between the initial and final, time-evolved state.
Note that zero loss (unit fidelity) does not necessarily imply that
the overall process is adiabatic, however,
a cyclic and completely adiabatic process should always yield $w_\mathrm{loss} = 0$.
Similar measures, but with an adiabatically evolved state as reference, have been used in Refs.~\cite{Karzig2015a,Amorim2015}.
In the simple two-Majorana braiding setup of Fig.~\ref{fig_1}, however, superpositions of states of different total parity are unphysical, and fidelities in a fixed parity sector lack information about the braid phase.
Further, in the full Hilbert space of the interacting Kitaev chain \eqref{eq:KitaevChain}, obtaining general analytic results, even under adiabatic evolution, appears impossible.

Our braiding protocol describes a cyclic evolution in the space of Hamiltonian parameters, where the Hamiltonian takes its original form at the end of the process, i.e., at time $t_\text{f} = 6T$.
Assuming adiabaticity, one can measure a Berry phase \cite{Berry1984} acquired during this cyclic evolution of the system, where the time-evolving state should always correspond to an eigenstate of the instantaneous Hamiltonian.
The fidelity after execution of the braid protocol, however, is not exactly unity (we find $w_\mathrm{loss} > 0$) because of non-adiabatic spectral losses due to the finite ramping time.
This in fact will manifest in a non-constant dynamical evolution of the acquired phase even after the system is evolved back to the initial Hamiltonian, cf. Fig.~\ref{fig_2}.
As a consequence, the Berry phase is not applicable to realistic, finite-time braiding protocols.
The unitary time-evolution of states under a continuous time-dependent
Hamiltonian, however, still describes a smooth curve in the Hilbert space.
Since quantum states that differ merely by a phase factor give rise to the same physics, one may then employ the gauge- and parametrization-invariant functional \cite{Samuel1988,Mukunda1993}
\begin{equation}
 \label{eq:geom_phase}
 \phi_\text{g}[C_0] = \arg\braket{\psi(t_0)}{\psi(t_\text{f})}
                    - \Im\int\limits_{t_0}^{t_\text{f}}\braket{\psi(t)}{\dot{\psi(t)}}\mathrm{d}t~,
\end{equation}
which measures the geometrical phase for the smooth open curve of normalized (unit) vectors in the Hilbert space
$\mathcal{C}_0=\{\ket{\psi(t)}\in \mathcal{N}_0\ |\ t\in[t_0,t_\text{f}]\subset \mathbb{R} \}\subset \mathcal{N}_0$.
Here $\mathcal{N}_0$ denotes the space of normalized vectors and $C_0$ is the projection of $\mathcal{C}_0$ to the so called ray space (or projective Hilbert space), where members that differ only by a phase factor are regarded as equivalent
\footnote{
An equivalence class of states in ray space is a projection operator {$|\psi \rangle \langle \psi|$} to the equivalence class represented by {$| \psi \rangle$}. This is a natural projection that maps each vector to the ray on which it lies.}.
We note that if the fidelity between the initial and final states is exactly one (zero loss), $C_0$ becomes a closed curve in ray space, and $\phi_\text{g}[C_0]$ corresponds to the Aharonov-Anandan geometrical phase \cite{Aharonov1987}.
If the process is also adiabatic, one recovers the conventional Berry phase \cite{Berry1984} (for discussion, cf. Ref.~\cite{Mukunda1993}).
In our time-dependent many-particle problem, the ramping (and hence braiding) protocol is accomplished numerically by considering the piece-wise constant, time-independent Hamiltonian in Eq.~\eqref{eq:time_evol}.
Therefore, $C_0$ is a continuous but generally not a smooth curve.
To determine the geometric phase numerically, we hence employ a discretized variant of the above functional \cite{Mukunda1993}
\begin{eqnarray}
 \label{eq:geom_phase_bargmann}
 \phi_\mathrm{g}[\tilde{C}_0] &=& \arg(\braket{\psi(t_0)}{\psi(t_\mathrm{f})})
  - \arg\left(\prod\limits_{j=0}^{N-1}
  \braket{\psi(t_j)}{\psi(t_{j+1})}\right)                     \nonumber \\
  &=& - \arg(\braket{\psi(t_0)}{\psi(t_1)}\braket{\psi(t_1)}{\psi(t_2)}\cdots \nonumber \\
  & & \qquad\cdots \braket{\psi(t_{N-1})}{\psi(t_\mathrm{f})}\braket{\psi(t_\mathrm{f})}{\psi(t_0)}),
\end{eqnarray}
which measures the geometric phase for the closed ${N+1}$-sided polygon of unit vectors in the Hilbert space
$\tilde{\mathcal{C}}_0=\left\{\ket{\psi(t_j)}\in \mathcal{N}_0\ |\ t_j = t_0 + j \Delta t,~j=0,\dots,N\right\}$, with$N = t_\mathrm{f}/\Delta t$.
The closed polygon $\tilde{C}_0$ parametrizes the braid protocol and corresponds to a projection of time-ordered set $\tilde{\mathcal{C}}_0$ onto ray space.
This expression is also referred to as the Bargmann vertex formula for the geometric phase. 
In the continuous limit $\Delta t \rightarrow 0$ in Eq. \eqref{eq:time_evol}, the expression in Eq.~\eqref{eq:geom_phase_bargmann} reduces to the continuous-time version in Eq.~\eqref{eq:geom_phase}.
Note also that the final expression above is cyclically symmetric.
In terms of a physical interpretation, the first and second term in Eqs.\eqref{eq:geom_phase}/\eqref{eq:geom_phase_bargmann} correspond to the complete and the local, dynamic parts of the acquired phase, respectively.
We thus explicitly exclude purely dynamic contributions, for example those generated by a finite Majorana hybridization \cite{Alicea2011,Clarke2011,Cheng2011}, from our considerations.

\paragraph{Exact diagonalization.}
The initial states $\ket{e}$ and $\ket{o}$ are chosen as the ground states of $H_{\mathrm{Y}}(t_0)$ in even and odd total-parity sectors, obtained through Davidson exact diagonalization \cite{Sadkane1999}.
The braiding is then carried out iteratively by the Arnoldi approximation of the product of matrix exponentials \eqref{eq:time_evol} with a time-evolving state vector \cite{MolerLoan2003}.
Since the time-dependent Hamiltonian conserves fermion parity, the simulation for each total-parity sector can be carried out separately.
Further, while strictly speaking the geometric phase is only defined after the braid process is carried out completely and a closed loop in parameter space is accomplished, it is convenient to define the exchange phase at arbitrary time $t$ of the braiding process,
\begin{equation}\label{eq:phasediff}
 \Phi(t) = \phi^{o}_\mathrm{g}(t) - \phi^{e}_\mathrm{g}(t)\, .
\end{equation}
Here, $\phi^{e}_\mathrm{g}(t)$ and $\phi^{o}_\mathrm{g}(t)$ are the individual phases acquired during the time evolution of states $\ket{e}$ and $\ket{o}$.
We note
${\ket{e}~\rightarrow~\e^{i\phi^{e}_\mathrm{g}(t)}\ket{e}}$ and 
${\ket{o}~\rightarrow~\e^{i\phi^{o}_\mathrm{g}(t)}\ket{o}}$,
and in the following drop the sub-script ``$\mathrm{g}$'' for brevity.

\begin{figure}
 \includegraphics[width=\columnwidth]{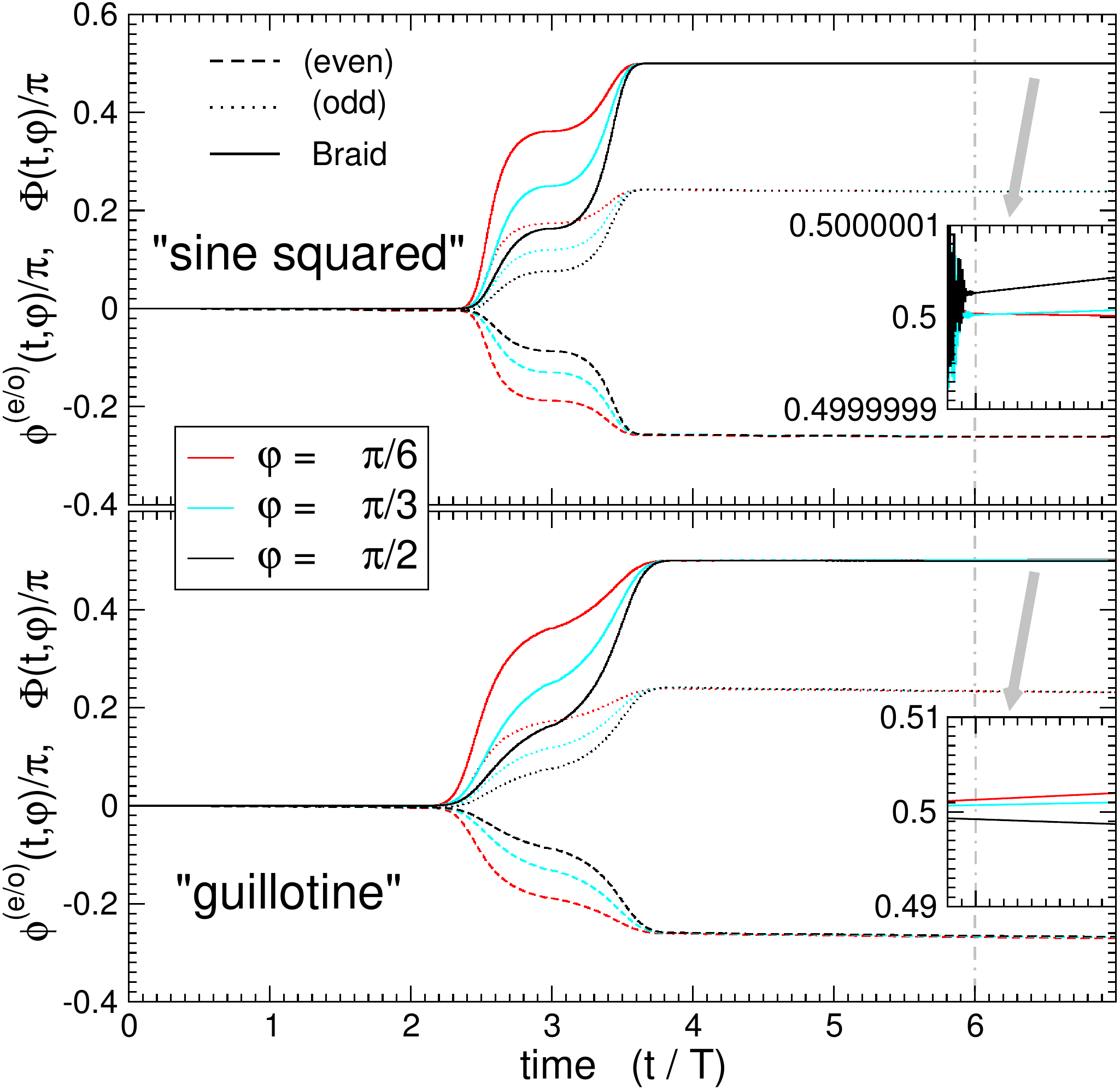}
 \caption{\label{fig_2}
          Time evolution of the acquired exchange phases $\phi^{e/o}(t,\varphi)$ (dashed/dotted lines),
          and the phase difference $\Phi(t,\varphi)$ (solid lines)
          during the braiding protocol for sine-squared (top) and guillotine ramp (bottom).
          We consider a Y-junction with equal-size legs $\ell = 5$, for triplet-pairing phases $\varphi=\pi/6,~\pi/3,~\pi/2$ and amplitude $\Delta=1$.
          The braid protocol takes a total time $t_\mathrm{f} = 6T$, where we choose a time period $T=750$ for each ramp-up/down step, with $\alpha=0.025$.
          After complete execution of the braiding exchange, upon reaching $t = t_\mathrm{f}$ (dash-dot vertical line),
          an extra time-evolution of duration $T$ is carried out with fixed Hamiltonian $H(t_\mathrm{f})$.
          Insets show the residual dynamic evolution of the exchange phase $\Phi(t)$, for $t\in [5.8T, 7T]$, zoomed in on $y$-scale.
         }
\end{figure} 

We now are equipped to study the symmetric Y-junction with equal-size legs $\ell = 5$,
where qualitatively similar results were found for leg sizes $\ell\leqslant 4$.
At the ideal Kitaev point $\Delta = J = 1$, the lowest-energy (initial) states in the even- and odd-parity sector correspond to $\ket{0}$ and $\ket{1}$ states of the topological qubit, i.e., they differ in terms of the Majorana sector only.
After the exchange ``braiding'' of Majorana modes, in general, both states will acquire different geometrical phases $\phi^\mathrm{e}$ and $\phi^\mathrm{o}$.
In an ideal case the accumulated many-body geometrical phase for $\ket{0}$ and $\ket{1}$ (and hence for $\ket{e}$ and $\ket{o}$) will only differ by a non-trivial phase stemming from the protected braid exchange of Majoranas $\gamma_\mathrm{L}$ and $\gamma_\mathrm{R}$.
In Fig.~\ref{fig_2} we show the time evolution of phases $\phi^\mathrm{e}(t,\varphi)$ and $\phi^\mathrm{o}(t,\varphi)$, as well as the phase difference $\Phi(t,\varphi)$ in Eq.~\eqref{eq:phasediff}, for sine-squared and guillotine ramp protocols and several superconducting pairing phases.
Here, after finishing the braiding protocol at time $t_\mathrm{f} = 6T$, we continue the time evolution for yet another period $T$.
In the fully adiabatic case, the geometrical phases in each parity sector -- and hence the difference between them -- should remain unchanged for times $t>t_\mathrm{f}$, i.e., when the braiding exchange has been executed completely.
In our case, however, the phases keep evolving approximately linearly
in time (insets in Fig.~\ref{fig_2}). We wish to emphasize that this phase evolution is \emph{not} due to purely dynamical phases as, e.g., caused by finite Majorana hybridization \cite{Clarke2011,Cheng2011}, which are explicitly excluded in Eq.~\eqref{eq:geom_phase_bargmann}. Instead, we can rationalize the continued non-trivial evolution by observing that the fidelity of the obtained states at time $t=t_\mathrm{f}$ is not exactly one, or equivalently the loss as defined in Eq.~$\eqref{eq:lost_fidelity}$ is non-zero, cf. Fig.~\ref{fig_4} and discussion.
Since in the adiabatic (zero-loss) limit the system should have reached a state equivalent to the initial state but with the Majoranas exchanged, this implies that instead, the system no longer resides in an eigenstate of its Hamiltonian for $t\geqslant t_\mathrm{f}$.
Such non-adiabatic spectral losses hence correspond to a leakage of wave function weight out of the Majorana, and into the excited-state sector.
Consequently, the state  at $t_\mathrm{F}\geqslant t_\mathrm{f}$ -- when projected to the (initial) qubit Hilbert space $\{\ket{0},\ket{1}\}$ -- will ``dephase'' with time.
This finding emphasizes the danger of non-adiabatic errors and
excitations even in the most basic braiding experiments, and cannot be
alleviated by just increasing the system size~\cite{Pedrocchi2015a,Knapp2016}.
Since there is only a weak dependence on $\varphi$ as shown in
Fig.~\ref{fig_2}, we will constrain ourselves to the symmetric case
$\varphi = \pi/3$ in the following. Note that for the finite
system size considered, this corresponds to the biggest time-evolving protective gap among all other geometric configurations of Y-junctions.

\begin{figure}
 \includegraphics[width=\columnwidth]{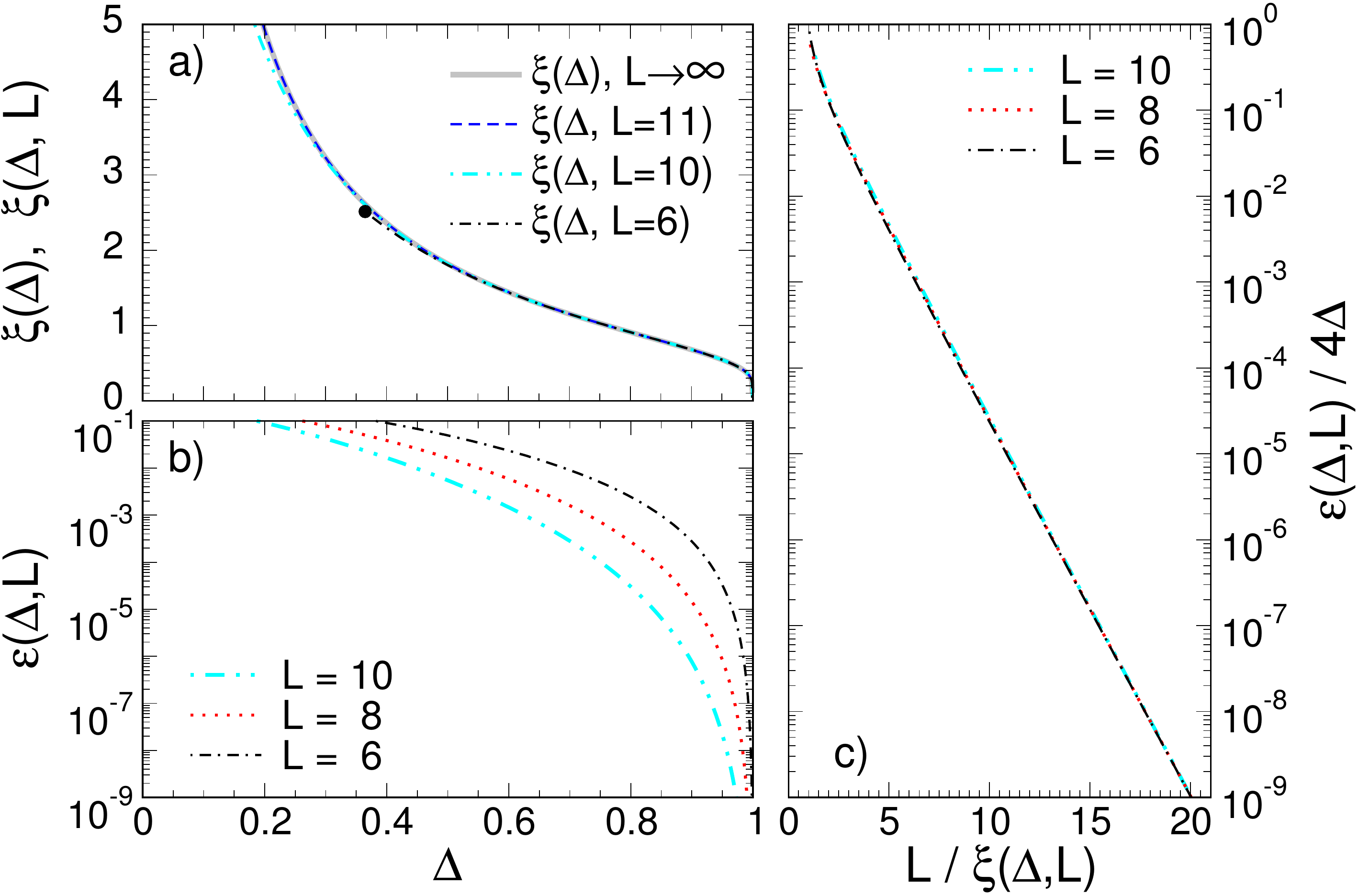}
 \caption{\label{fig_3}
          a)~MBS localization length for infinite, $\xi(\Delta)$, and finite-size Kitaev chains, $\xi(\Delta,L)$, as a function of pairing amplitude $\Delta$. Results for the infinitely long (bold gray curve) and finite Kitaev chains match closely already for modest $L = 11$ (dashed curve) and $L=6$ (dash-dotted curve).
          b)~MBS hybridization energy $\varepsilon(\Delta,L)$ for finite-size Kitaev chains.
          c)~Semi-log plot of scaled MBS hybridization energy $\varepsilon(\Delta,L)/4\Delta$ vs. scaled inverse localization length $L/\xi(\Delta, L)$.
          For comparable $L$ and $\xi(\Delta, L)$ there is a clear deviation from the exponential dependence (upper left corner).
         }
\end{figure}

\begin{figure*}
 \includegraphics[width=\textwidth]{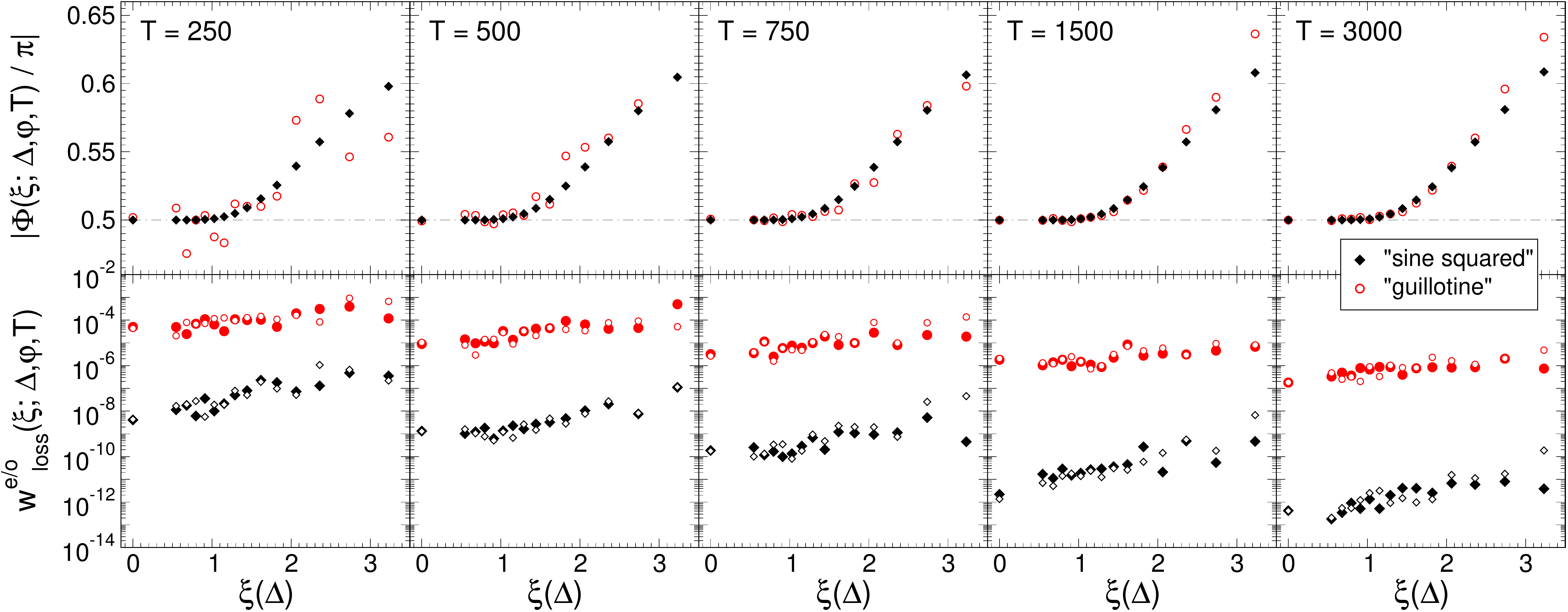}
 \caption{\label{fig_4}
          Exchange phase $\Phi(\xi;\Delta,\varphi,T)$ (top row) and
          loss of fidelity $w^{e/o}_\mathrm{loss}(\xi;\Delta, \varphi,T)$ (bottom row)
          vs.~MBS localization length $\xi(\Delta)$, for sine-squared (black diamonds) and guillotine ramp (red circles).
          We consider a symmetric Y-junction ($\varphi = \pi/3$) with equal-size legs $\ell = 5$.
          Plots from left to right correspond to execution of the braid protocol with ramping time periods of
          $T=250,~500,~750,~1500$, and $3000$, respectively, with $\alpha=0.025$.
          Data points (left-to-right) for each plot correspond to pairing amplitudes $\Delta=1.0,\dots,~0.3$ in steps $0.05$, which translates to increasing MBS localization length $\xi(\Delta)$ from left to right.
          The loss of fidelity (bottom row) in even and odd total-parity sectors is shown by filled and open symbols, respectively.
         }
\end{figure*}

\paragraph{Braiding errors.}
There are several sources of errors that destroy the perfect quantization of the braiding phase during the non-adiabatic, unitary time-evolution of the system.
One is seen as a reduced fidelity due to the spectral losses, cf. Fig.~\ref{fig_2} and the discussion above, which are expected to be larger for smaller protective gap ($\sim \Delta$), lower ramp time $T$, and rougher ramp functions.
Further, the condition of exponential localization of the two Majorana zero modes is invalidated when the MBS localization length becomes comparable with the chain size, see Fig.~\ref{fig_3}.
In this case, the hybridization between MBSs is not exponentially small, and the braiding phase encodes dynamic effects stemming from this hybridization.
Both the minimal protective gap as well as Majorana localization length depend on the $p$-wave pairing amplitude $\Delta$.
Here, $\xi_\infty(\Delta)=1/\ln\sqrt{(J+|\Delta|)/(J-|\Delta|)}$ corresponds to the localization length in the infinitely
long Kitaev chain with $\mu=0$ where exact results are easily obtained. In Fig.~\ref{fig_3}a we show $\xi_\infty(\Delta)$ and the finite-length $\xi(\Delta,L)$ as a function of $\Delta$.
The latter is determined by fitting the weights of the fermionic representation of the MBSs to an exponentially decaying function modulated by Friedel oscillations with period $\pi$~\cite{Stoudenmire2011}.
$L=11$ and $L=6$ chain sizes correspond to the maximal and minimal length of the topological regions during the braiding in the symmetric Y-junction with equal-size legs $\ell = 5$, cf. Fig.~\ref{fig_1}.
Perfectly localized MBSs, $\xi\to 0$, are obtained at $J=\Delta=1$.
For $\Delta<1$, MBSs acquire a finite width that increases with decreasing values of $\Delta$, and the localization length $\xi(\Delta)$ diverges for vanishing $\Delta$.
We note that the numerical result $\xi(\Delta, L=6)$ starts do deviate from $\xi_\infty(\Delta)$ at $\Delta \lesssim 0.5$, indicating that MBSs are not isolated anymore and the independent-particle picture \cite{Ivanov2001,Alicea2011} breaks down.

In Fig.~\ref{fig_4} we show the braiding phase and fidelity losses as a function of the Majorana localization length $\xi(\Delta)$ for the two ramp protocols and various step times $T$.
Data in Fig.~\ref{fig_4} (bottom row) shows that fidelity losses \eqref{eq:lost_fidelity} reduce by orders of magnitude upon increasing the ramping time period $T$ from $250$ to $3000$.
In addition, losses can be reduced significantly (by a power of $2$) upon employing the smoother sine-squared ramp function (first derivative is continuous) instead of the simpler guillotine one (first derivative jumps), agreeing with the power-law scaling identified by Knapp et al.~\cite{Knapp2016}.
Similarly, non-adiabatic effects measured directly in the exchange phase \eqref{eq:phasediff}, see Fig.~\ref{fig_4}a (top), which can be assessed by the strength of phase-variations as function of the braid parameters (i.e. the sensitivity of the braid phase against small deviations in the protocol), become smaller. In the adiabatic limit for large $T$, these fluctuations become negligible and one obtains smooth curves $\Phi(\xi,\Delta,\dots)$.
The braiding phase is stable and closest to $\pi/2$, independent of
$T$, for the ideal parameter set $\Delta = J = 1$.
As expected, for non-ideal values $\Delta < 1$, stable braid phases
are obtained only for larger times $T$. There are, however, significant deviations from the desired $\pi/2$ braiding
phase starting from $\xi > 1$ (corresponding to $\Delta< 0.7$), which persist largely independent of the protocol time $T$ or smoothness of the ramp function.
These can be identified with a break-down of the independent-particle picture of Majorana braiding, where ideal braid statistics do no longer apply.
The braid phase in this sense is more sensitive to finite Majorana wave-function overlaps than spectral features like the MBS hybridization, or the predicted localization length in Fig.~\ref{fig_3}.
We emphasize that even though the braid process appears quasi-adiabatic for $T\geqslant 500$ (or $T\geqslant 1500$ for guillotine ramp) and spectral losses even for large $\xi$ are substantially smaller than under quicker braid routines, only the exchange phase itself encodes this dramatic breakdown of braiding.
Matching the Majorana wave-function weights or checking for adiabacity alone may thus be too loose a criterion for successful Majorana braid protocols \cite{Karzig2015a,Amorim2015}.

\begin{figure}
 \includegraphics[width=0.965\columnwidth]{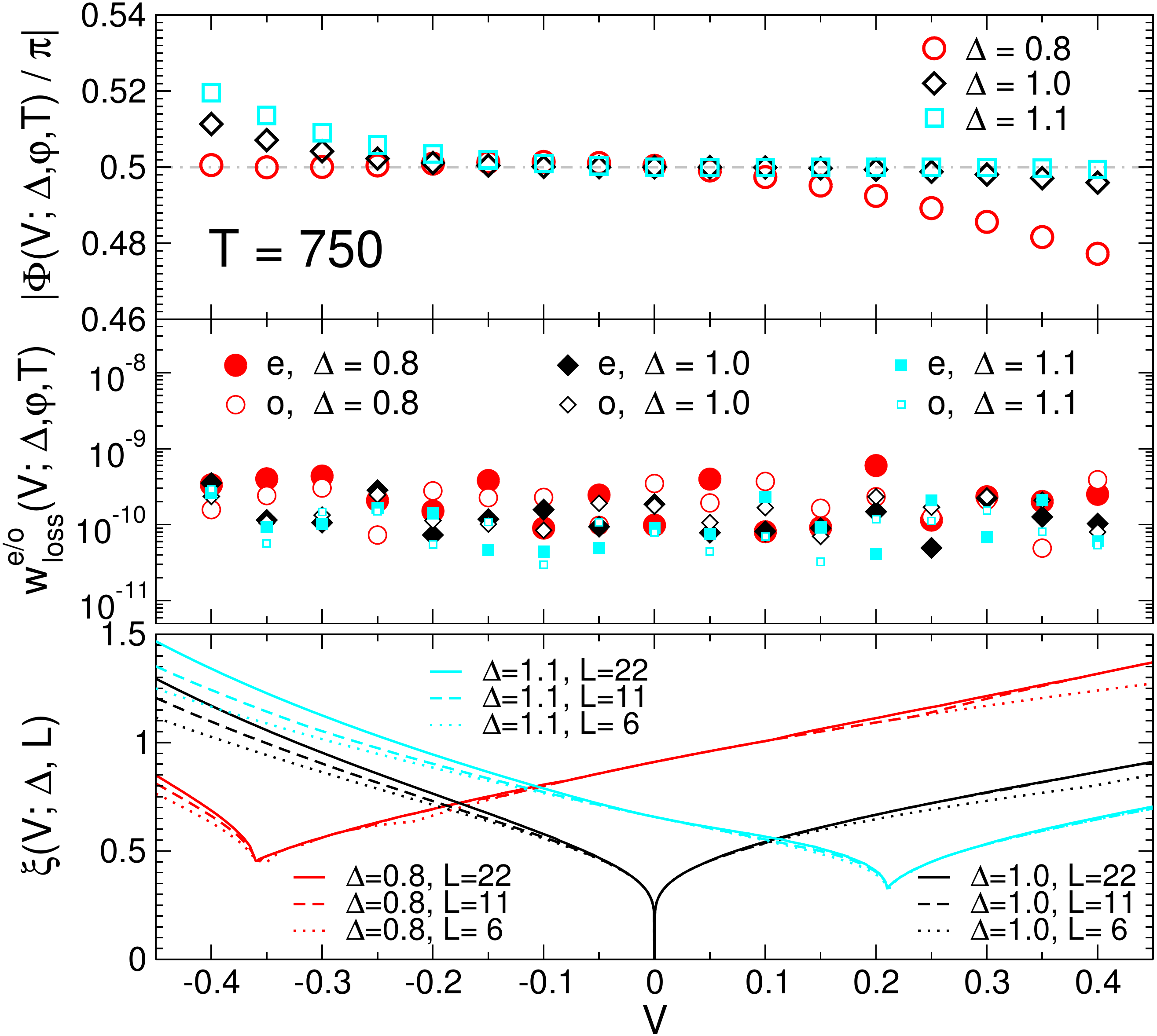}
 \caption{\label{fig_5}
          Exchange phase $\Phi(V;\Delta,\varphi,T)$ (upper plot)
          and loss of fidelity $w^{e/o}_\mathrm{loss}(V; \Delta,\varphi,T)$ (middle plot)
          vs. interaction strength $V$ for pairing amplitudes $\Delta = 1.0$ (black diamonds), $\Delta =0.8$ (red circles) and $\Delta = 1.1$ (cyan squares).
          As in Fig.\ref{fig_4}, the braid protocol is applied to a symmetric Y-junction ($\varphi=\pi/3$) with equal-size legs $\ell = 5$, using 
          the sine-squared ramp with time period $T=750$ and $\alpha=0.025$.
          The bottom plot shows the MBS localization length $\xi(V; \Delta, L)$ vs. interaction strength $V$ for a Kitaev chain with $L=22,~11,~6$ sites (solid, dashed, and dotted lines). Again, we consider pairing amplitudes $\Delta=1.0$ (black), $\Delta=0.8$ (red) and $\Delta = 1.1$ (cyan).
         }
\end{figure}

\paragraph{Interactions.}
Finally, we investigate the influence of a nearest neighbor 
interaction $V\neq 0$, see Eqs.~\eqref{eq:KitaevChain} and \eqref{eq:y_junction}, on the braiding operation.
In general, interactions in Majorana systems modify the localization
length and change the bulk energy gap
\cite{Loss2011,Sela2011,Stoudenmire2011,ThomaleRachelSchmitteckert2013}. Large
values of interaction strength $V$ of either sign eliminate the topological phase.
Here, we only consider values $V$ for which the topological phase is not destroyed at $\mu=0$, and where $\mu=4$ is still large enough to drive the corresponding junction leg into the trivial phase.
In Fig.~\ref{fig_5}, we plot the braiding phase $\Phi(V;\Delta,\varphi,T)$ and loss of fidelity $w^{e/o}_\mathrm{loss}(V;\Delta,\varphi,T)$ as a function of $V$. We consider a symmetric Y-junction with equal-size legs $\ell = 5$, now with pairing strengths $\Delta =1.0,~0.8$, and $1.1$.
We only show results for the sine-squared ramp function with $T=750$, which in the non-interacting case was found to be sufficient to ensure close-to-adiabatic conditions, cf. Fig.~\ref{fig_4}.
The Majorana localization length $\xi(V;\Delta,L)$ is depicted for
Kitaev chains of up to $L=22$ sites, and obtained by fitting the weights of the free fermionic representation of the MBS~\cite{Stoudenmire2011}.
For the ideal parameter set $\Delta=J=1$, the Majorana localization
length increases with either increasing repulsive or attractive interaction $V$.
This effect turns out to be larger for an attractive potential, where $\xi>1$ for $\vert V \vert >0.35$, and the braiding phase starts to deviate significantly from $\pi/2$.
For $\Delta=0.8$, the minimum of $\xi(V;\Delta=0.8,L)$ is shifted to
$V\approx -0.35$, where for the given $T$ one also observes a
stabilization of the correct braiding phase.
For large $\Delta = 1.1$, the situation is reversed since then repulsive interactions lead to a reduction of the effective superconducting gap, pushing the system closer to the ideal point $\Delta_\mathrm{eff} = J$.
The fidelity losses mainly depend on the braid velocity $\sim \ell/T$, and hence remain of similar magnitude for moderate $V$.
As a weak trend, we find that small attractive or repulsive interactions play a stabilizing role for the braiding operation, depending on the relative size of the bare couplings $J$ and $\Delta$. 
Again, these effects appear not to be systematically encoded in spectral losses alone, which one can investigate by tracking the Majorana wave-function weights \cite{Amorim2015}, but rather follow directly the behavior of the MBS localization length $\xi$, cf. both Figs. \ref{fig_4} and \ref{fig_5}.

\paragraph{Discussion and conclusions.}
Previous studies of Majorana braiding have predominantly focused on effective models of the low-energy (Majorana) sector \cite{Alicea2011,Clarke2011},
introducing non-adiabatic effects and other sources of error on
an effective but not microscopic level \cite{Cheng2011,Karzig2013,Scheurer2013}. 
Alternatively, the fidelity of the dynamically evolved state was considered with respect to some ideal, adiabatic reference state \cite{Karzig2015a,Karzig2015b,Amorim2015}.
Resulting estimates for the minimal protocol time for near-adiabatic
time-evolution in Majorana systems were obtained by many groups~\cite{Karzig2013,Scheurer2013,Karzig2015a,Karzig2015b,Amorim2015}. 
As an overarching implication, until recently, it was assumed that by just executing the braiding manipulations sufficiently slowly, one could recover the ideal operation completely.
This result, however, disregards the generation and propagation of
non-adiabatic excitations \cite{Knapp2016}, which in turn directly affect the geometric braiding phase \cite{Pedrocchi2015a}.

In our work, instead of further extending previous studies to more
complex setups and protocols \cite{Karzig2015a,Karzig2015b,Aasen2016},
we have revisited the most elementary case of a single MBS braiding operation in a Y-junction of three Kitaev chains \cite{Alicea2011}.
We considered the full Hilbert space, including the quasiparticle background beyond the low-energy Majorana sector, and investigated several error sources that destroy the perfect braiding phase during the non-adiabatic, unitary time-evolution of the system.
By adopting the Bargmann vertex formula for the geometric phase, customarily used in the characterization of geometric quantum gate operations \cite{Mukunda1993}, we identified explicitly the MBS braiding phase, and analyzed its behavior during the braid.
We then accounted for both spectral losses (through the fidelity) and
direct braiding phase errors (by calculating the geometric phase
evolution explicitly) in a numerically exact framework, in order to understand in detail how non-adiabacity, the Majorana mode localization, and interactions affect the braiding operation.

We find that due to the induced non-adiabatic leakage of wavefunction weight out of the ground state into excited states of the final Hamiltonian, the acquired braiding phase does not remain constant and evolves linearly in time even after the braiding process is finished, cf. Fig.~\ref{fig_2}.
Consequently, the final (qubit) state will ``dephase'' with time, even in the zero-temperature case considered here.
While spectral losses can be reduced by orders of magnitude through
increasing the ramping time period $T$ or considering smoother ramp
functions, cf.~Fig.~\ref{fig_4}, it is clear that in extended
protocols such as necessary for serious quantum computations, non-adiabatic errors will accumulate.
Our results, based on a numerically exact treatment of the full
Hilbert space, thus support and extend the findings of Pedrocchi and
DiVincenzo~\cite{Pedrocchi2015a} and Knapp et al.~\cite{Knapp2016}
that employed an effective description of the low-energy Majorana sector (incl. quasiparticle background via baths/dissipation).

By varying the superconducting wire bulk gap, we additionally resolved how the localization length of individual Majorana zero modes affects the braiding operation, beyond simple dynamical errors induced through a finite MBS hybridization \cite{Cheng2011,Amorim2015}.
For the studied Y-junction configuration in Fig.~\ref{fig_1}, a MBS localization length $\xi>1$ already causes significant deviations from the perfect braiding phase $\pi/2$, since the independent-particle picture of Majorana braiding \cite{Ivanov2001,Kitaev2001,Alicea2011} breaks down.
We emphasize that this finite-size breakdown occurs independent of
adiabacity, cf.~Fig.~\ref{fig_4}. It implies that measures of
adiabacity -- such as the fidelity or Majorana wave function weights
alone -- may suggest overoptimistic efficiencies of the braiding operation.
At the same time, the behavior in Fig.~\ref{fig_4} shows how an extraction of the geometric phase \eqref{eq:geom_phase_bargmann} can yield valuable information about the anyonic statistics encoded in the many-body states of a complex system.
Finally, nearest neighbor interaction in the wires is found to predominantly affect the braiding operation through its impact on the wire bulk gap.
Depending on the non-interacting gap versus hopping strength, for weak
attractive (repulsive) interaction, the braiding operation is more
stable due to a change of the effective superconducting gap and the reduction of the MBS localization length, cf. Refs. \cite{Loss2011,Stoudenmire2011,Sela2011}.

In summary, finite operation times and non-ideal parameter settings in (simple) Majorana braiding schemes pose serious constraints on the accuracy of braiding operations.
We found that even for the ideal case of a closed, zero-temperature Kitaev chain system, non-adiabatic errors and finite Majorana hybridization can become bottlenecks for the fidelity and feasibility of (extended) Majorana braiding routines towards quantum computation.
An accumulation of non-adiabatic errors has to be avoided, likely leading to unfavorable finite-time scaling if no explicit quasiparticle relaxation mechanism is included.
The optimal engineering of dissipation out of the excited-state sector, such as trapping of quasiparticles, as well as measurement-based (dissipative) topological quantum computation schemes \cite{Plugge2017,Vijay2016,StationQ2016} may thus behave favorably in comparison to conventional adiabatic-manipulation schemes \cite{Alicea2011,Hyart2013,Aasen2016}, cf. Ref.~\cite{Knapp2016}.
Similarly, finite Majorana overlaps have to be avoided to high accuracy, since these directly affect the braiding statistics encoded in the low-energy sector of the system -- in fact, the braid statistics become ``non-Majorana'', cf. Fig.~\ref{fig_4} and discussion.
As an order of magnitude comparison, we note that current-date
nanowire and iron-chain architectures \cite{Albrecht2016,Deng2016,Zhang2016,Pawlak2016,Feldman2017} find
localization vs. device lengths of $\xi/L \simeq 1 ... 5$, where our numerics indicate that assuming individual Majoranas (in terms of their braiding statistics) might be too optimistic.
Advancing towards longer nanowire devices or a reduction of the MBS localization length, also for the sake of measurement-based qubit experiments \cite{Plugge2017,Vijay2016,StationQ2016}, is thus clearly desirable.

Finally, an analysis of braiding and computation protocols in extended
Majorana wire networks
\cite{Landau2016,Plugge2016,Plugge2017,Vijay2016,StationQ2016} will be the subject to future work, using more sophisticated techniques applicable to larger system sizes.
Clearly, these measurement-based schemes for topological quantum computation should face the same detailed numerical investigation and scrutiny as the direct braiding and adiabatic-manipulation schemes.

\begin{acknowledgments}
We thank A. Akhmerov, R. Egger, S. Frolov, and R. Lutchyn for fruitful discussions.  This work was supported by DFG-SFB 1170 and ERC-StG-Thomale-TOPOLECTRICS-336012.
M.S. acknowledges support by the Rustaveli national science foundation through the grand no. FR/265/6-100/14.
\end{acknowledgments}

\bibliography{msmrp_paper}

\end{document}